\documentclass[conference]{IEEEtran}%
\usepackage{amsmath}
\usepackage{amsfonts}
\usepackage{amssymb}
\usepackage{float}
\usepackage{url}
\usepackage{mathrsfs}
\usepackage{graphicx}%
\IEEEoverridecommandlockouts

\newtheorem{definition}{Definition}

\newtheorem{notation}{Convention}

\def\textsc#1{\uppercase{#1}}
\DeclareSymbolFont{rsfscript}{OMS}{rsfs}{m}{n}
\DeclareSymbolFontAlphabet{\mathrsfs}{rsfscript}
\newcommand{\oldcal}[1]{}
\let\oldcal=\mathcal
\renewcommand{\mathcal}{\mathrsfs}
\begin{document}

\title{An Automated Analysis of the Security \\of Quantum Key Distribution}%

\def\rajaack{\thanks{R. Nagarajan is supported by EPSRC grant GR/S34090
and the EU Sixth Framework Programme (Project SecoQC: \textit
{Development of a Global Network for Secure Communication based on Quantum Cryptography}%
).}}
\def\nickack{\thanks{N. Papanikolaou is supported by a Postgraduate Fellowship
by the Department of Computer Science, University of Warwick.}}
\def\garyack{\thanks{G. Bowen is supported by EPSRC grant GR/S92816.}}
\author{\authorblockN{Rajagopal Nagarajan\rajaack
\  and\\ Nikolaos Papanikolaou\nickack}
\authorblockA{Department of Computer Science\\
University of Warwick\\
United Kingdom\\
Email: \{biju,nikos\}@dcs.warwick.ac.uk}
\and\authorblockN{Garry Bowen\garyack}
\authorblockA{Centre for Quantum Computation\\
University of Cambridge\\
United Kingdom\\
Email: gab30@cam.ac.uk}
\and\authorblockN{Simon Gay}
\authorblockA{Department of Computing Science\\
University of Glasgow\\
United Kingdom\\
Email: simon@dcs.gla.ac.uk} }%
%

\maketitle
%

\begin{abstract}%

This paper discusses the use of computer--aided verification as a practical
means for analysing quantum information systems; specifically, the BB84
protocol for quantum key distribution is examined using this method. This
protocol has been shown to be unconditionally secure against all attacks in an
information--theoretic setting, but the relevant security proof requires a
thorough understanding of the formalism of quantum mechanics and is not easily
adaptable to practical scenarios. Our approach is based on \emph{probabilistic
model--checking;} we have used the PRISM model--checker to show that, as the
number of qubits transmitted in BB84 is increased, the equivocation of the
eavesdropper with respect to the channel decreases exponentially. We have also
shown that the probability of detecting the presence of an eavesdropper
increases exponentially with the number of qubits. The results presented here
are a testament to the effectiveness of the model--checking approach for
systems where analytical solutions may not be possible or plausible.%

\end{abstract}%

\section{Introduction}

That quantum--mechanical phenomena can be effectively exploited for the
storage, manipulation and exchange of information is now a widely recognised
fact. The whole field of quantum information poses new challenges for the
information theory community and involves several novel applications,
especially with respect to cryptology.

Recent interest in quantum cryptography has been stimulated by the fact that
quantum algorithms, such as Shor's algorithms for integer factorization and
discrete logarithm \cite{Shor94}, threaten the security of classical
cryptosystems. A range of quantum cryptographic protocols for key
distribution, bit commitment, oblivious transfer and other problems
\cite{Brassard91} have been extensively studied. Furthermore, the
implementation of quantum cryptographic protocols has turned out to be
significantly easier than the implementation of quantum algorithms: although
practical quantum computers are still some way in the future, quantum
cryptography has already been demonstrated in non-laboratory settings
\cite{PoppeA:praqkd} and is well on the way to becoming an important technology.

Quantum cryptographic protocols are designed with the intention that their
security is guaranteed by the laws of quantum physics. Naturally it is
necessary to prove, for any given protocol, that this is indeed the case. The
most notable result in this area is Mayers' proof \cite{Mayers01} of the
unconditional security of the quantum key distribution protocol
\textquotedblleft BB84\textquotedblright\ \cite{bb84-orig}. This proof
guarantees the security of BB84 in the presence of an attacker who can perform
any operation allowed by quantum physics; hence the security of the protocol
will not be compromised by future developments in quantum computing.

Mayers' result, and others of the same kind
\cite{lochau-proof,newproof,Mayers96}, are extremely important contributions
to the study of quantum cryptography. However, a mathematical proof of
security of a \emph{protocol} does not in itself guarantee the security of an
implemented \emph{system} which relies on the protocol. Experience of
classical cryptography has shown that, during the progression from an
idealised protocol to an implementation, many security weaknesses can arise.
For example: the system might not correctly implement the desired protocol;
there might be security flaws which only appear at the implementation level
and which are not visible at the level of abstraction used in proofs; problems
can also arise at boundaries between systems and between components which have
different execution models or data representations. We therefore argue that it
is worth analysing quantum cryptographic systems at a level of detail which is
closer to a practical implementation.

Computer scientists have developed a range of techniques and tools for the
analysis and verification of communication systems and protocols. Those
particularly relevant to security analysis are surveyed by Ryan \emph{et al.}
\cite{RyanP:modasp}. This approach has two key features. The first is the use
of formal languages to precisely specify the behaviour of the system and the
properties which it is meant to satisfy. The second is the use of automated
software tools to either verify that a system satisfies a specification or to
discover flaws. These features provide a high degree of confidence in the
validity of systems, and the ability to analyse variations and modifications
of a system very easily.

In this paper we present the results of applying the above methodology to the
BB84 quantum key distribution protocol. We have carried out an analysis using
PRISM\footnote{See
\url{http://www.cs.bham.ac.uk/~dxp/prism}%
.}, a probabilistic model-checking system. Our results confirm the properties
which arise from Mayers' security proof; more significantly, they demonstrate
the effectiveness of the model-checking approach and the ease with which
parameters of the system can be varied.

Our model could easily be adapted to describe variations and related
protocols, such as \textquotedblleft B92\textquotedblright\ and Ekert's
protocol (\cite{gruska,nielsenchuang}\ describe these protocols in detail).
Also, our model can be modified to account for implementation--level concerns,
such as imperfections in photon sources, channels, and detectors.

\section{Quantum Key Distribution \newline and Security
Criteria\label{bb84-sec}}

The objective of \emph{key distribution} is to enable two communicating
parties, Alice and Bob, to agree on a common secret key $\vec{k}\in
\{0,1\}^{N},\,N>0$, without sharing any information initially. Once a common
secret key has been established, Alice and Bob can use a symmetric
cryptosystem to exchange messages privately. In a classical (i.e.
non--quantum) setting, it is quite impossible to perform key distribution
securely unless assumptions are made about the enemy's computational power
\cite{gruska}.

The use of quantum channels, which cannot be tapped or monitored without
causing a noticeable disturbance, makes unconditionally secure key
distribution possible. The presence of an enemy is made manifest to the users
of such channels through an unusually high error rate. We will now describe
the BB84 scheme for quantum key distribution, which uses polarised photons as
information carriers.

BB84 assumes that the two legitimate users are linked by two specific
channels, which the enemy also has access to:

\begin{enumerate}
\item a classical, possibly public channel, which may be passively monitored
but not tampered with by the enemy;

\item a quantum channel which may be tampered with by an enemy. By its very
nature, this channel prevents passive monitoring.
\end{enumerate}

\noindent The first phase of BB84 involves transmissions over the quantum
channel, while the second phase takes place over the classical channel.

\begin{notation}
The pair of quantum states $\{\left\vert 0\right\rangle ,\left\vert
1\right\rangle \}$ is the \emph{rectilinear basis} of the Hilbert space
$\mathcal{H}_{2}$, and is denoted by $\boxplus$.
\end{notation}

\begin{notation}
The pair of quantum states $\{\frac{1}{\sqrt{2}}\left(  \left\vert
0\right\rangle +\left\vert 1\right\rangle \right)  ,\frac{1}{\sqrt{2}}\left(
\left\vert 0\right\rangle -\left\vert 1\right\rangle \right)  \}$ is the
\emph{diagonal basis} of the Hilbert space $\mathcal{H}_{2}$, and is denoted
by $\boxtimes$.
\end{notation}

\begin{definition}
The encoding function $f_{\mathrm{BB84}}:D\times B\mapsto\mathcal{H}_{2}$
where $D=\{0,1\},$ $B=\{\boxplus,\boxtimes\}$ is defined as follows:
\begin{align}
f_{\mathrm{BB84}}(0,\boxplus)  &  =\left\vert 0\right\rangle \\
f_{\mathrm{BB84}}(1,\boxplus)  &  =\left\vert 1\right\rangle \\
f_{\mathrm{BB84}}(0,\boxtimes)  &  =\frac{1}{\sqrt{2}}\left(  \left\vert
0\right\rangle +\left\vert 1\right\rangle \right) \\
f_{\mathrm{BB84}}(1,\boxtimes)  &  =\frac{1}{\sqrt{2}}\left(  \left\vert
0\right\rangle -\left\vert 1\right\rangle \right)
\end{align}

\end{definition}

The BB84 protocol can be summarised as follows:

\begin{enumerate}
\item First Phase (Quantum Transmissions)

\begin{enumerate}
\item Alice generates a random string of bits $\vec{d}\in\{0,1\}^{n}$, and a
random string of bases $\vec{b}\in\{\boxplus,\boxtimes\}^{n}$, where $n>N$.

\item Alice places a photon in quantum state $\left\vert \psi_{i}\right\rangle
=f_{\mathrm{BB84}}(d_{i},b_{i})$ for each bit $d_{i}$ in $\vec{d}$ and $b_{i}$
in $\vec{b}$, and sends it to Bob over the quantum channel.

\item Bob measures each $\left\vert \psi_{i}\right\rangle $ received, with
respect to either $\boxplus$ or $\boxtimes$, chosen at random. Bob's
measurements produce a string $\overrightarrow{d^{\prime}}\in\{0,1\}^{n}$,
while his choices of bases form $\overrightarrow{b^{\prime}}\in\{\boxplus
,\boxtimes\}^{n}$.
\end{enumerate}

\item Second Phase (Public Discussion)

\begin{enumerate}
\item For each bit $d_{i}$ in $\vec{d}$:

\begin{enumerate}
\item Alice sends the value of $b_{i}$ to Bob over the classical channel.

\item Bob responds by stating whether he used the same basis for measurement.
If $b_{i}^{\prime}\neq b_{i}$, both $d_{i}$ and $d_{i}^{\prime}$ are discarded.
\end{enumerate}

\item Alice chooses a subset of the remaining bits in $\vec{d}$ and discloses
their values to Bob over the classical channel. If the result of Bob's
measurements for any of these bits do not match the values disclosed,
eavesdropping is detected and communication is aborted.

\item The common secret key, $\overrightarrow{k}\in\{0,1\}^{N}$, is the string
of bits remaining in $\vec{d}$ once the bits disclosed in step \textit{2b)}
are removed.
\end{enumerate}
\end{enumerate}

There are two points to note in order to understand BB84 properly. Firstly,
measuring with the incorrect basis yields a random result, as predicted by
quantum theory. Thus, if Bob chooses the $\boxtimes$ basis to measure a photon
in state $\left\vert 1\right\rangle $, the classical outcome will be either 0
or 1 with equal probability; if the $\boxplus$ basis was chosen instead, the
classical outcome would be 1 with certainty. Secondly, in step \textit{2b)} of
the protocol, Alice and Bob perform a test for eavesdropping. The idea is
that, wherever Alice and Bob's bases are identical (i.e. $b_{i}^{\prime}%
=b_{i}$), the corresponding bits should match (i.e.~$d_{i}^{\prime}=d_{i}$).
If not, an external disturbance has occurred, and on a noiseless channel this
can only be attributed to the presence of an eavesdropper. For more
information, the reader is referred to \cite{gruska, nielsenchuang}.

We turn now to the formal security requirements for BB84. Among other things,
a protocol such as BB84 must ensure that an enemy's presence is always made
manifest to the legitimate users and that, if a key does result from the
procedure, it is unpredictable and common to both users. But most importantly,
the protocol must ensure \emph{privacy:} an enemy must never be able to obtain
the value of the key. Moreover, even if an enemy is able to obtain a certain
quantity of information $\vec{v}$ by trying to monitor the classical channel,
that quantity has to be minimal; meanwhile, the enemy's uncertainty about the
key, $\mathrm{H}(\vec{k}|\vec{v})$, must be maximised.

\begin{definition}
The conditional entropy of the key $\vec{k}$ (of length $N$) given the view
$\vec{v}$ is defined as:
\[
\mathrm{H}_{N}(\vec{k}|\vec{v})=-\frac{1}{\Pr\{N\}}%
{\displaystyle\sum\limits_{k}}
{\displaystyle\sum\limits_{v}}
\Pr\{k,v\}\log\left(  \Pr\{k|v\}\right)
\]

\end{definition}

Such requirements are usually expressed in terms of \emph{security
parameters}. For quantum key distribution, the security parameters are
conventionally written $n$ and $\vec{\epsilon}$. The parameter $n$ is the
number of quantum states transmitted, while $\vec{\epsilon}$ denotes
collectively the tolerated error rate, the number of bits used to test for
eavesdropping, and related quantities \cite{Mayers01}. We use the parameter
$n$ instead of the key length $N$, as these are assumed to be linearly
related. For instance, the value of $\mathrm{H}(\vec{k}|\vec{v})$\ is some
function of $n$ and $\vec{\epsilon}$: $\mathrm{H}(\vec{k}|\vec{v}%
)=\varphi(n,\vec{\epsilon})$. The proof \cite{Mayers01} stipulates that
$\mathrm{H}(\vec{k}|\vec{v})$\ should be exponentially small in $n$ and
$\vec{\epsilon}$. Formally,
\begin{align}
\varphi(n,\vec{\epsilon})  &  \leqslant c\cdot\mathrm{e}^{-gn}%
\label{decaying-expo}\\
\lim_{n\rightarrow\infty}\varphi(n,\vec{\epsilon})  &  =0
\end{align}
noting that the choice of $n$ over $N$ as the parameter only changes the value
of the constant $g$, and not the functional relationship. \noindent We will
demonstrate later for BB84 that, the probability that an enemy succeeds in
obtaining more than $\frac{n}{2}$ key bits correctly is a function of the form
(\ref{decaying-expo}).

Mayers' security proof of BB84\ formalises the notion of privacy by defining a
quantum key distribution protocol as \textquotedblleft$f$%
--private,\textquotedblright\ if, for every strategy adopted by an enemy, the
average of the quantity $N-H(\vec{k}|\vec{v})$ is less than or equal to some
constant $f$. This definition of privacy merely requires the key to be
uniformly distributed, when the key length $N$ is known. A more conventional
privacy definition would have required that the mutual information
$\mathrm{I}(\vec{k},\vec{v})$ be less than or equal to $\kappa$, but this is
not entirely satisfactory \cite{Mayers01}.

\section{Model Checking Techniques\newline and the PRISM Tool}

The theoretical proof of BB84's security is a significant and valuable result.
However, to prove a similar result for a different scheme or cryptographic
task is far from trivial and is likely to involve new, ever more specialised
derivations. A more flexible approach to analysing the security of quantum
cryptographic protocols is clearly desirable. Manufacturers of commercial
quantum cryptographic systems \cite{muller:793}, for instance, require
efficient and rigorous methods for design and testing.\ A suitable approach
should allow for modelling implementation--level details and even minor
protocol variations with relative ease. We believe that \emph{model--checking}
is such an approach, and we will demonstrate its application to BB84.

Model--checking is an automated technique for verifying a finite--state system
against a given temporal specification \cite{doronpeled}. Using a specialised
software tool (called a \emph{model--checker}), a system implementor can
mechanically prove that the system satisfies a certain set of requirements. To
do this, an abstract model, denoted $\sigma$, is built and expressed in a
description language; also, the desired behaviour of the system is expressed
as a set of temporal formulae, $\Phi_{i}$. The model and the formulae are then
fed into the model--checker, whose built--in algorithms determine conclusively
whether $\sigma$ satisfies the properties defined by the $\Phi_{i}$ (i.e.
whether $\sigma\models\Phi_{i}$ for each property $\Phi_{i}$%
).\ Model--checking should not be confounded with computer--based simulation
techniques, which do not involve an exhaustive search of all possibilities.

For systems which exhibit probabilistic behaviour, a variation of this
technique is used; a \emph{probabilistic model--checker,} such as PRISM
\cite{prism-manual},\ computes the probability
\begin{equation}
\Pr\{\sigma\models\Phi_{i}\} \label{el-probabilidad}%
\end{equation}
for given $\sigma$ and $\Phi_{i}$. PRISM\ models are represented by
probabilistic transition systems, and are written in a simple guarded--command
programming language. System properties for PRISM models are written in
Probabilistic Computation Tree Logic (PCTL).

PRISM allows models to be parameterised: $\sigma=\sigma(u_{1},\ldots,u_{k})$.
Thus the probability (\ref{el-probabilidad}) may be computed for different
values of $u_{1},\ldots,u_{k}$; this is termed an \emph{experiment}. By
varying one parameter at a time, it is possible to produce a meaningful plot
of the variation of (\ref{el-probabilidad}).

\section{Analysis of BB84 using PRISM}

We have built a model of BB84 for use with PRISM. It is not possible to
present the source code for this model here, due to space limitations;
however, the full source code is available online\footnote{See
{\raggedright \url{http://go.warwick.ac.uk/nikos/research/publications/index.html}
}.}, and is discussed extensively in \cite{nikosthesis}.

A system description in PRISM is a computer file containing module
definitions, each module representing a component of the system. In our
description of BB84, there is a module corresponding to each party involved in
the protocol and a module representing the quantum channel. Each module has a
set of local variables and a sequence of actions to perform; an action
typically takes one of the following two forms:
\begin{align}
& [s]\text{ \ }g\text{ }\rightarrow\text{ }(v_{1}:=\mathrm{val}_{1}%
);\label{firstcase}\\
& [s]\text{ \ }g\text{ }\rightarrow0.5:(v_{1}:=\mathrm{val}_{1})+0.5:(v_{1}%
:=\mathrm{val}_{2});\label{secondcase}%
\end{align}

\noindent In (\ref{firstcase}), the variable $v_{1}$ is assigned the value
$\mathrm{val}_{1}$; in (\ref{secondcase}), $v_{1}$ is assigned either the
value $\mathrm{val}_{1}$ or $\mathrm{val}_{2}$\ with equal probability. Part
of the expressive power of PRISM comes from the ability to specify arbitrary
probabilities for actions; for example, one could model a bias in Alice's
choice of polarisation basis, in BB84, with an action such as:%

\begin{align}
\lbrack\mathit{choosebasis}]\text{ }\mathbf{true}\rightarrow0  &
.7:(\mathit{al\_basis}:=\boxplus)\label{varyingalicebasis}\\
+0  &  .3:(\mathit{al\_basis}:=\boxtimes);\nonumber
\end{align}

\noindent In this example, Alice is biased towards choosing the rectilinear
basis. Knowledge of this syntax is sufficient for an understanding of the
PRISM description of BB84. In what follows, we will discuss the properties
which we have been able to investigate.

As discussed in section \ref{bb84-sec}, there are two security requirements
for BB84 of interest:

\begin{enumerate}
\item \emph{an enemy's presence must not go unnoticed;} if the legitimate
users know that an enemy is trying to eavesdrop, they can agree to use privacy
amplification techniques [20] and/or temporarily abort the key establishment process.

\item \emph{any quantity of valid information which the enemy is able to
obtain through eavesdropping must be minimal.}
\end{enumerate}

We can use our model of BB84, denoted henceforth by $\sigma_{\mathrm{BB84}}$,
to compute the probability
\begin{equation}
\Pr\{\sigma_{\mathrm{BB84}}\models\Phi_{i}\}
\end{equation}

\noindent where $\Phi_{i}$ is a given PCTL property--formula. Therefore, in
order to verify that BB84 satisfies the security requirements just mentioned,
we have to reformulate these requirements in terms of probability.

Firstly, we should be able to compute exactly what the probability of
detecting an enemy is. In our PRISM model, we can vary $n$, the number of
photons transmitted in a trial of BB84, and so this probability is a function
of $n$. Let us write the probability of detecting an enemy as%

\begin{equation}
P_{\det}(n)=\Pr\{\sigma_{\mathrm{BB84}}\models\Phi_{\det}\} \label{dudakis}%
\end{equation}

In (\ref{dudakis}), $\Phi_{\det}$ represents the PCTL formula whose boolean
value is $\mathtt{true}$ when an enemy is detected. Before we give the
definition of $\Phi_{\det}$, we should state the random event $\mathcal{E}$
that occurs when an enemy is detected; this will allow us to write $P_{\det
}(n)$ as a classical probability $\Pr(\mathcal{E})$.

In BB84, an enemy, Eve, is detected as a result of the disturbance inevitably
caused by some of her measurements. Just as Bob, Eve does not know which
polarisation bases were used to encode the bits in Alice's original bit
string. Eve has to make a random choice of basis, denoted $b_{i}^{\prime
\prime}$, which may or may not match Alice's original choice, $b_{i}$. If
$b_{i}^{\prime\prime}=b_{i}$, Eve is guaranteed to measure the $i$-th photon
correctly; otherwise, quantum theory predicts that her measurement result will
only be correct with probability 0.5.

In a so--called \emph{intercept--resend attack,} Eve receives each photon on
the quantum channel, measures it with her basis $b_{i}^{\prime\prime}$,
obtaining bit value $d_{i}^{\prime\prime}$, and then transmits to Bob a new
photon, which represents $d_{i}^{\prime\prime}$ in the $b_{i}^{\prime\prime}$
basis. If Eve's basis choice is incorrect, her presence is bound to be
detected. But for detection to occur, Bob must choose the correct basis for
his measurement. Whenever Bob obtains an incorrect bit value despite having
used the correct basis, this is because an enemy has caused a disturbance.
Note that we are assuming a perfect quantum channel here; an imperfect channel
would produce noise, causing additional disturbances.

So, to summarise, an enemy's presence is made manifest as soon as the
following event occurs:%

\[
(b_{i}^{\prime\prime}\neq b_{i})\wedge(b_{i}^{\prime}=b_{i})\text{ for some
}i\leqslant n
\]

\noindent or equivalently, as soon as:%

\begin{equation}
\mathcal{E}\equiv(b_{i}^{\prime}=b_{i})\wedge(d_{i}^{\prime}\neq d_{i})\text{
for some }i\leqslant n
\end{equation}

Therefore, the probability of detecting an enemy's presence in BB84 may be written:%

\begin{align*}
P_{\det}(n)  &  =\Pr\{\mathcal{E}\}\\
&  =\Pr\{(b_{i}^{\prime}=b_{i})\wedge(d_{i}^{\prime}\neq d_{i})\text{ for some
}i\leqslant n\}
\end{align*}

\noindent The corresponding PCTL formula for PRISM is:%

\[
\Phi_{\det}\equiv\{\mathtt{true}\text{ }\mathcal{%
\oldcal{U}%
}\text{ }(b_{i}^{\prime}=b_{i})\wedge(d_{i}^{\prime}\neq d_{i})\}
\]

\noindent The PRISM model of BB84 uses elaborate variable names, e.g.
$\mathtt{bob\_basis}$ instead of $b_{i}^{\prime}$, and $\mathtt{alice\_bit}$
instead of $d_{i}$.

The value of $P_{\det}(n)$ for $5\leqslant n\leqslant30$ has been calculated
with PRISM, which computes (\ref{dudakis}); the result is shown in Figure 1.%

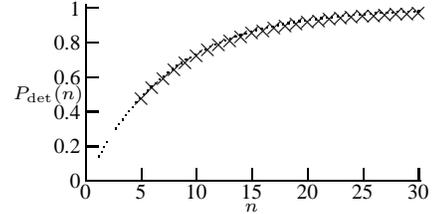
\begin{figure}[hb]
\begin{center}
\setlength{\unitlength}{0.1pt} \ifx\plotpoint\undefined\newsavebox{\plotpoint
}\fi\sbox{\plotpoint}{\rule[-0.200pt]{0.400pt}{0.400pt}}
\begin{picture}(1500,750)(0,0)
\sbox{\plotpoint}{\rule[-0.200pt]{0.400pt}{0.400pt}}
\put(181.0,123.0){\rule[-0.200pt]{4.818pt}{0.400pt}}
\put(161,123){\makebox(0,0)[r]{\footnotesize0}}
\put(181.0,254.0){\rule[-0.200pt]{4.818pt}{0.400pt}}
\put(161,254){\makebox(0,0)[r]{\footnotesize{0.2}}}
\put(181.0,385.0){\rule[-0.200pt]{4.818pt}{0.400pt}}
\put(161,385){\makebox(0,0)[r]{\footnotesize{0.4}}}
\put(181.0,515.0){\rule[-0.200pt]{4.818pt}{0.400pt}}
\put(161,515){\makebox(0,0)[r]{\footnotesize{0.6}}}
\put(181.0,646.0){\rule[-0.200pt]{4.818pt}{0.400pt}}
\put(161,646){\makebox(0,0)[r]{\footnotesize{0.8}}}
\put(181.0,777.0){\rule[-0.200pt]{4.818pt}{0.400pt}}
\put(161,777){\makebox(0,0)[r]{\footnotesize{1}}}
\put(181.0,123.0){\rule[-0.200pt]{0.400pt}{4.818pt}}
\put(181,82){\makebox(0,0){\footnotesize{0}}}
\put(391.0,123.0){\rule[-0.200pt]{0.400pt}{4.818pt}}
\put(391,82){\makebox(0,0){\footnotesize{5}}}
\put(600.0,123.0){\rule[-0.200pt]{0.400pt}{4.818pt}}
\put(600,82){\makebox(0,0){\footnotesize{10}}}
\put(810.0,123.0){\rule[-0.200pt]{0.400pt}{4.818pt}}
\put(810,82){\makebox(0,0){\footnotesize{15}}}
\put(1020.0,123.0){\rule[-0.200pt]{0.400pt}{4.818pt}}
\put(1020,82){\makebox(0,0){\footnotesize{20}}}
\put(1229.0,123.0){\rule[-0.200pt]{0.400pt}{4.818pt}}
\put(1229,82){\makebox(0,0){\footnotesize{25}}}
\put(1439.0,123.0){\rule[-0.200pt]{0.400pt}{4.818pt}}
\put(1439,82){\makebox(0,0){\footnotesize{30}}}
\put(40,450){\makebox(0,0){$\scriptstyle{P_{\mathrm{det}}(n)}$}}
\put(810,21){\makebox(0,0){$\scriptstyle{n}$}}
\put(391,442){\raisebox{-.8pt}{\makebox(0,0){$\times$}}}
\put(433,483){\raisebox{-.8pt}{\makebox(0,0){$\times$}}}
\put(475,520){\raisebox{-.8pt}{\makebox(0,0){$\times$}}}
\put(516,552){\raisebox{-.8pt}{\makebox(0,0){$\times$}}}
\put(558,580){\raisebox{-.8pt}{\makebox(0,0){$\times$}}}
\put(600,605){\raisebox{-.8pt}{\makebox(0,0){$\times$}}}
\put(642,626){\raisebox{-.8pt}{\makebox(0,0){$\times$}}}
\put(684,645){\raisebox{-.8pt}{\makebox(0,0){$\times$}}}
\put(726,662){\raisebox{-.8pt}{\makebox(0,0){$\times$}}}
\put(768,676){\raisebox{-.8pt}{\makebox(0,0){$\times$}}}
\put(810,689){\raisebox{-.8pt}{\makebox(0,0){$\times$}}}
\put(852,700){\raisebox{-.8pt}{\makebox(0,0){$\times$}}}
\put(894,709){\raisebox{-.8pt}{\makebox(0,0){$\times$}}}
\put(936,718){\raisebox{-.8pt}{\makebox(0,0){$\times$}}}
\put(978,725){\raisebox{-.8pt}{\makebox(0,0){$\times$}}}
\put(1020,732){\raisebox{-.8pt}{\makebox(0,0){$\times$}}}
\put(1062,737){\raisebox{-.8pt}{\makebox(0,0){$\times$}}}
\put(1104,742){\raisebox{-.8pt}{\makebox(0,0){$\times$}}}
\put(1145,747){\raisebox{-.8pt}{\makebox(0,0){$\times$}}}
\put(1187,750){\raisebox{-.8pt}{\makebox(0,0){$\times$}}}
\put(1229,754){\raisebox{-.8pt}{\makebox(0,0){$\times$}}}
\put(1271,757){\raisebox{-.8pt}{\makebox(0,0){$\times$}}}
\put(1313,759){\raisebox{-.8pt}{\makebox(0,0){$\times$}}}
\put(1355,761){\raisebox{-.8pt}{\makebox(0,0){$\times$}}}
\put(1397,763){\raisebox{-.8pt}{\makebox(0,0){$\times$}}}
\put(1439,765){\raisebox{-.8pt}{\makebox(0,0){$\times$}}}
\put(228.69,215.14){\usebox{\plotpoint}}
\put(239.13,233.07){\usebox{\plotpoint}}
\put(249.57,251.01){\usebox{\plotpoint}}
\put(260.16,268.85){\usebox{\plotpoint}}
\put(293.87,321.20){\usebox{\plotpoint}}
\put(305.91,338.10){\usebox{\plotpoint}}
\put(318.44,354.65){\usebox{\plotpoint}}
\put(330.54,371.51){\usebox{\plotpoint}}
\put(343.40,387.79){\usebox{\plotpoint}}
\put(356.89,403.56){\usebox{\plotpoint}}
\put(370.48,419.25){\usebox{\plotpoint}}
\put(384.00,435.00){\usebox{\plotpoint}}
\put(398.67,449.67){\usebox{\plotpoint}}
\put(413.21,464.48){\usebox{\plotpoint}}
\put(427.73,479.29){\usebox{\plotpoint}}
\put(443.30,493.02){\usebox{\plotpoint}}
\put(459.14,506.43){\usebox{\plotpoint}}
\put(474.62,520.24){\usebox{\plotpoint}}
\put(491.07,532.90){\usebox{\plotpoint}}
\put(507.26,545.88){\usebox{\plotpoint}}
\put(524.06,558.04){\usebox{\plotpoint}}
\put(541.57,569.17){\usebox{\plotpoint}}
\put(558.74,580.83){\usebox{\plotpoint}}
\put(576.77,591.09){\usebox{\plotpoint}}
\put(594.54,601.81){\usebox{\plotpoint}}
\put(612.71,611.84){\usebox{\plotpoint}}
\put(631.27,621.08){\usebox{\plotpoint}}
\put(649.77,630.43){\usebox{\plotpoint}}
\put(668.97,638.29){\usebox{\plotpoint}}
\put(688.00,646.58){\usebox{\plotpoint}}
\put(707.11,654.57){\usebox{\plotpoint}}
\put(726.54,661.81){\usebox{\plotpoint}}
\put(746.05,668.86){\usebox{\plotpoint}}
\put(765.89,674.97){\usebox{\plotpoint}}
\put(785.64,681.35){\usebox{\plotpoint}}
\put(805.46,687.49){\usebox{\plotpoint}}
\put(825.40,693.17){\usebox{\plotpoint}}
\put(845.61,697.90){\usebox{\plotpoint}}
\put(865.80,702.72){\usebox{\plotpoint}}
\put(886.02,707.39){\usebox{\plotpoint}}
\put(906.21,712.19){\usebox{\plotpoint}}
\put(926.60,715.98){\usebox{\plotpoint}}
\put(946.97,719.92){\usebox{\plotpoint}}
\put(967.36,723.75){\usebox{\plotpoint}}
\put(987.86,726.98){\usebox{\plotpoint}}
\put(1008.36,730.21){\usebox{\plotpoint}}
\put(1028.86,733.48){\usebox{\plotpoint}}
\put(1049.48,735.69){\usebox{\plotpoint}}
\put(1069.97,738.99){\usebox{\plotpoint}}
\put(1090.66,740.59){\usebox{\plotpoint}}
\put(1111.24,743.19){\usebox{\plotpoint}}
\put(1131.83,745.67){\usebox{\plotpoint}}
\put(1152.51,747.46){\usebox{\plotpoint}}
\put(1173.20,749.09){\usebox{\plotpoint}}
\put(1193.89,750.68){\usebox{\plotpoint}}
\put(1214.58,752.35){\usebox{\plotpoint}}
\put(1235.28,753.94){\usebox{\plotpoint}}
\put(1255.96,755.61){\usebox{\plotpoint}}
\put(1276.66,757.20){\usebox{\plotpoint}}
\put(1297.38,758.00){\usebox{\plotpoint}}
\put(1318.08,759.47){\usebox{\plotpoint}}
\put(1338.78,761.00){\usebox{\plotpoint}}
\put(1359.50,761.73){\usebox{\plotpoint}}
\put(1380.23,762.40){\usebox{\plotpoint}}
\put(1400.93,763.99){\usebox{\plotpoint}}
\put(1421.66,764.64){\usebox{\plotpoint}}
\put(1439,765){\usebox{\plotpoint}}
\put(181.0,123.0){\rule[-0.200pt]{127pt}{0.400pt}}     \put(181.0,123.0){\rule
[-0.200pt]{0.400pt}{67pt}}
\end{picture}
\end{center}
\caption
{The probability that Eve is detected in the BB84 Protocol while performing an intercept--resend attack,
as a function of the security parameter $n$. The crosses indicate data points produced by PRISM, while the
dotted curve is a non--linear least--squares fit to these points.}
\end{figure}%

The first requirement for BB84, namely that it should be possible to detect an
enemy's presence, clearly is satisfied. As we can see from Figure 1, as the
number of photons transmitted is increased, the probability of detection tends
toward 1, i.e. we conclude that%

\[
\lim_{n\rightarrow\infty}P_{\det}(n)=1
\]

We will now consider the second security requirement. Let $\mathcal{C}_{i}$
denote the event in which Eve measures the $i$-th photon transmitted
correctly. The probability that Eve measures all photons correctly, and hence
is able to obtain the secret key, is the product%

\[
P_{\mathrm{all}}=%
{\displaystyle\prod\limits_{0<i\leqslant n}}
\Pr\{\mathcal{C}_{i}\}=\Pr\{\mathcal{C}_{1}\}\Pr\{\mathcal{C}_{2}\}\cdots
\Pr\{\mathcal{C}_{n}\}
\]

\noindent We will examine the variation of a quantity proportional to
$P_{\mathrm{all}}$, namely the probability $P_{>1/2}(n)$ that Eve measures
\emph{more than half} the photons transmitted correctly.

According to the second security requirement for BB84, the amount of valid
information obtained by an enemy must be minimised; we will investigate the
variation of the probability%

\[
P_{>1/2}(n)=\Pr\{\sigma_{\mathrm{BB84}}\models\Phi_{>1/2}\}
\]

\noindent as a function of the number of photons transmitted. We expect this
quantity to grow smaller and smaller with $n$.

The PRISM model of BB84 includes a counter variable, \texttt{nc}, whose value
is the number of times that Eve makes a correct measurement. The formula
$\Phi_{>1/2}$ may be written in terms of this variable:
\[
\Phi_{>1/2}=\left\{  \mathtt{true}\text{ }\mathcal{%
\oldcal{U}%
}\text{ }\left(  \mathtt{nc}>\frac{n}{2}\right)  \right\}
\]

Given $\sigma_{\mathrm{BB84}}$ and $\Phi_{>1/2}$, PRISM produces the plot
shown in Figure 2; it can be seen from the figure that $P_{>1/2}(n)$ decays
exponentially with $n$.%

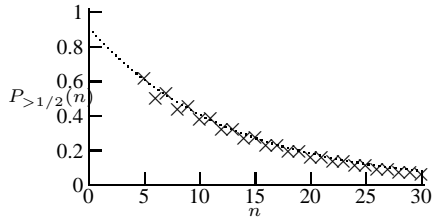
\begin{figure}[ht]
\begin{center}
\setlength{\unitlength}{0.1pt}\ifx\plotpoint\undefined\newsavebox{\plotpoint
}\fi\sbox{\plotpoint}{\rule[-0.200pt]{0.400pt}{0.400pt}}
\begin{picture}(1500,750)(0,0)
\sbox{\plotpoint}{\rule[-0.200pt]{0.400pt}{0.400pt}}
\put(181.0,123.0){\rule[-0.200pt]{4.818pt}{0.400pt}}
\put(161,123){\makebox(0,0)[r]{\footnotesize0}}
\put(181.0,254.0){\rule[-0.200pt]{4.818pt}{0.400pt}}
\put(161,254){\makebox(0,0)[r]{\footnotesize{0.2}}}
\put(181.0,385.0){\rule[-0.200pt]{4.818pt}{0.400pt}}
\put(161,385){\makebox(0,0)[r]{\footnotesize{0.4}}}
\put(181.0,515.0){\rule[-0.200pt]{4.818pt}{0.400pt}}
\put(161,515){\makebox(0,0)[r]{\footnotesize{0.6}}}
\put(181.0,646.0){\rule[-0.200pt]{4.818pt}{0.400pt}}
\put(161,646){\makebox(0,0)[r]{\footnotesize{0.8}}}
\put(181.0,777.0){\rule[-0.200pt]{4.818pt}{0.400pt}}
\put(161,777){\makebox(0,0)[r]{\footnotesize{1}}}
\put(181.0,123.0){\rule[-0.200pt]{0.400pt}{4.818pt}}
\put(181,82){\makebox(0,0){\footnotesize{0}}}
\put(391.0,123.0){\rule[-0.200pt]{0.400pt}{4.818pt}}
\put(391,82){\makebox(0,0){\footnotesize{5}}}
\put(600.0,123.0){\rule[-0.200pt]{0.400pt}{4.818pt}}
\put(600,82){\makebox(0,0){\footnotesize{10}}}
\put(810.0,123.0){\rule[-0.200pt]{0.400pt}{4.818pt}}
\put(810,82){\makebox(0,0){\footnotesize{15}}}
\put(1020.0,123.0){\rule[-0.200pt]{0.400pt}{4.818pt}}
\put(1020,82){\makebox(0,0){\footnotesize{20}}}
\put(1229.0,123.0){\rule[-0.200pt]{0.400pt}{4.818pt}}
\put(1229,82){\makebox(0,0){\footnotesize{25}}}
\put(1439.0,123.0){\rule[-0.200pt]{0.400pt}{4.818pt}}
\put(1439,82){\makebox(0,0){\footnotesize{30}}}
\put(40,450){\makebox(0,0){$\scriptstyle{P_{>1/2}(n)}$}}
\put(810,21){\makebox(0,0){$\scriptstyle{n}$}}
\put(391,534){\raisebox{-.8pt}{\makebox(0,0){$\times$}}}
\put(433,458){\raisebox{-.8pt}{\makebox(0,0){$\times$}}}
\put(475,479){\raisebox{-.8pt}{\makebox(0,0){$\times$}}}
\put(516,418){\raisebox{-.8pt}{\makebox(0,0){$\times$}}}
\put(558,428){\raisebox{-.8pt}{\makebox(0,0){$\times$}}}
\put(600,378){\raisebox{-.8pt}{\makebox(0,0){$\times$}}}
\put(642,383){\raisebox{-.8pt}{\makebox(0,0){$\times$}}}
\put(684,341){\raisebox{-.8pt}{\makebox(0,0){$\times$}}}
\put(726,344){\raisebox{-.8pt}{\makebox(0,0){$\times$}}}
\put(768,309){\raisebox{-.8pt}{\makebox(0,0){$\times$}}}
\put(810,310){\raisebox{-.8pt}{\makebox(0,0){$\times$}}}
\put(852,281){\raisebox{-.8pt}{\makebox(0,0){$\times$}}}
\put(894,281){\raisebox{-.8pt}{\makebox(0,0){$\times$}}}
\put(936,257){\raisebox{-.8pt}{\makebox(0,0){$\times$}}}
\put(978,257){\raisebox{-.8pt}{\makebox(0,0){$\times$}}}
\put(1020,236){\raisebox{-.8pt}{\makebox(0,0){$\times$}}}
\put(1062,237){\raisebox{-.8pt}{\makebox(0,0){$\times$}}}
\put(1104,219){\raisebox{-.8pt}{\makebox(0,0){$\times$}}}
\put(1145,219){\raisebox{-.8pt}{\makebox(0,0){$\times$}}}
\put(1187,204){\raisebox{-.8pt}{\makebox(0,0){$\times$}}}
\put(1229,204){\raisebox{-.8pt}{\makebox(0,0){$\times$}}}
\put(1271,192){\raisebox{-.8pt}{\makebox(0,0){$\times$}}}
\put(1313,192){\raisebox{-.8pt}{\makebox(0,0){$\times$}}}
\put(1355,181){\raisebox{-.8pt}{\makebox(0,0){$\times$}}}
\put(1397,181){\raisebox{-.8pt}{\makebox(0,0){$\times$}}}
\put(1439,172){\raisebox{-.8pt}{\makebox(0,0){$\times$}}}
\put(181,718){\usebox{\plotpoint}}
\put(181.00,718.00){\usebox{\plotpoint}}
\put(194.59,702.31){\usebox{\plotpoint}}
\put(208.28,686.72){\usebox{\plotpoint}}
\put(222.81,671.90){\usebox{\plotpoint}}
\put(237.12,656.88){\usebox{\plotpoint}}
\put(251.80,642.20){\usebox{\plotpoint}}
\put(266.47,627.53){\usebox{\plotpoint}}
\put(281.59,613.30){\usebox{\plotpoint}}
\put(296.42,598.80){\usebox{\plotpoint}}
\put(312.27,585.39){\usebox{\plotpoint}}
\put(327.87,571.71){\usebox{\plotpoint}}
\put(343.53,558.09){\usebox{\plotpoint}}
\put(359.39,544.70){\usebox{\plotpoint}}
\put(375.72,531.90){\usebox{\plotpoint}}
\put(392.20,519.32){\usebox{\plotpoint}}
\put(408.83,506.90){\usebox{\plotpoint}}
\put(425.52,494.57){\usebox{\plotpoint}}
\put(442.58,482.75){\usebox{\plotpoint}}
\put(460.06,471.58){\usebox{\plotpoint}}
\put(476.97,459.56){\usebox{\plotpoint}}
\put(494.64,448.68){\usebox{\plotpoint}}
\put(512.07,437.42){\usebox{\plotpoint}}
\put(530.14,427.22){\usebox{\plotpoint}}
\put(548.18,416.98){\usebox{\plotpoint}}
\put(566.22,406.73){\usebox{\plotpoint}}
\put(584.50,396.89){\usebox{\plotpoint}}
\put(602.96,387.41){\usebox{\plotpoint}}
\put(621.49,378.08){\usebox{\plotpoint}}
\put(640.15,369.01){\usebox{\plotpoint}}
\put(659.00,360.31){\usebox{\plotpoint}}
\put(677.86,351.64){\usebox{\plotpoint}}
\put(697.11,343.88){\usebox{\plotpoint}}
\put(716.10,335.54){\usebox{\plotpoint}}
\put(735.35,327.79){\usebox{\plotpoint}}
\put(754.73,320.34){\usebox{\plotpoint}}
\put(774.23,313.26){\usebox{\plotpoint}}
\put(793.73,306.16){\usebox{\plotpoint}}
\put(813.24,299.15){\usebox{\plotpoint}}
\put(832.98,292.78){\usebox{\plotpoint}}
\put(852.74,286.42){\usebox{\plotpoint}}
\put(872.56,280.29){\usebox{\plotpoint}}
\put(892.64,275.08){\usebox{\plotpoint}}
\put(912.40,268.72){\usebox{\plotpoint}}
\put(932.47,263.51){\usebox{\plotpoint}}
\put(952.41,257.83){\usebox{\plotpoint}}
\put(972.64,253.16){\usebox{\plotpoint}}
\put(992.81,248.30){\usebox{\plotpoint}}
\put(1013.03,243.61){\usebox{\plotpoint}}
\put(1033.20,238.72){\usebox{\plotpoint}}
\put(1053.43,234.06){\usebox{\plotpoint}}
\put(1073.80,230.12){\usebox{\plotpoint}}
\put(1094.18,226.28){\usebox{\plotpoint}}
\put(1114.50,222.08){\usebox{\plotpoint}}
\put(1134.81,217.88){\usebox{\plotpoint}}
\put(1155.31,214.62){\usebox{\plotpoint}}
\put(1175.76,211.13){\usebox{\plotpoint}}
\put(1196.14,207.29){\usebox{\plotpoint}}
\put(1216.63,203.98){\usebox{\plotpoint}}
\put(1237.16,200.90){\usebox{\plotpoint}}
\put(1257.76,198.50){\usebox{\plotpoint}}
\put(1278.27,195.34){\usebox{\plotpoint}}
\put(1298.76,192.04){\usebox{\plotpoint}}
\put(1319.39,189.86){\usebox{\plotpoint}}
\put(1340.00,187.54){\usebox{\plotpoint}}
\put(1360.52,184.38){\usebox{\plotpoint}}
\put(1381.18,182.52){\usebox{\plotpoint}}
\put(1401.76,179.94){\usebox{\plotpoint}}
\put(1422.37,177.61){\usebox{\plotpoint}}
\put(1439,176){\usebox{\plotpoint}}
\put(181.0,123.0){\rule[-0.200pt]{127pt}{0.400pt}}     \put(181.0,123.0){\rule
[-0.200pt]{0.400pt}{67pt}}
\end{picture}
\end{center}
\caption
{The probability that Eve, by performing an intercept--resend attack, makes more than $\frac
{n}{2}$
correct measurements in BB84, versus the security parameter $n$.}
\end{figure}%

Figures 1 and 2 each contain two superimposed plots: the data points marked
with crosses are actual values produced by PRISM, and the dotted curves are
nonlinear functions to which the data points have been fitted. We have used
the Levenberg--Marquardt nonlinear fitting algorithm to compute values
$c_{1},c_{2},c_{3}$ and $c_{4}$ such that:%

\begin{align*}
P_{\det}(n)  &  \approx1-c_{1}\exp[-c_{2}n]\\
P_{>1/2}(n)  &  \approx c_{3}\exp[-c_{4}n]
\end{align*}

In particular, the values obtained are (to three decimal places): $c_{1}=1$,
$c_{2}=0.134$, $c_{3}=0.909$, and $c_{4}=0.081$. It is evident that,
increasing the number of photons transmitted, or equivalently, the length of
the bit sequence generated by Alice, increases BB84's capability to avert an
enemy: the probability of detecting the enemy increases exponentially, while
the amount of valid information the enemy has about the key decreases exponentially.

These results are in agreement with Mayers' claim (see \cite{Mayers01}), that
\textquotedblleft in an information--theoretic setting, which is our case, a
quantity $f_{N}$ such as the amount of Shannon's information available to Eve
must decrease exponentially fast as $N$ increases.\textquotedblright%
\ Remember, we have assumed that the number of transmissions, $n$, is linearly
related to $N$.

Variations in the protocol can be accommodated easily by modifying the PRISM
model. For example, in \cite{ardehali} a bias in Alice's choice of basis is
introduced, and this can be described by a PRISM action such as
(\ref{varyingalicebasis}). This influences the performance of BB84; it alters
the variation of both $P_{\det}(n)$ and $P_{>1/2}(n)$. It is also possible to
vary \emph{a posteriori} probabilities with PRISM, such as the probability
that, for any given transmission, the enemy's choice of measurement basis
matches Alice's original choice. This probability is not usually taken into
consideration in manual proofs, and is likely to be useful for modelling more
sophisticated eavesdropping attacks.

It should be noted that the results presented here are not as general as
Mayers'. For instance, we have assumed that a noiseless channel is being used,
and we have only considered a finite number of cases (namely, where
$5\leqslant n\leqslant30$). Related techniques from computer science, which
are better suited for a full proof of unconditional security, do exist; the
most appropriate of these is \emph{automated theorem proving}
\cite{doronpeled}; we will leave this for future work. This technique is not
restricted to finite scenarios, and can provide the generality needed for a
more extensive analysis.

\section{Conclusions}

In this paper we have analysed the security of the BB84 protocol for quantum
key distribution by applying \emph{formal verification techniques,} which are
well--established in theoretical computer science. In particular, an automated
model-checking system, PRISM, was used to obtain results which corroborate
Mayers' unconditional security proof of the protocol. Compared to manual
proofs of security, our approach offers several advantages. Firstly, it is
easily adapted to cater for other quantum protocols. It also allows us to
analyse composite systems, which include both classical and
quantum--mechanical components. Finally, we are not only able to model
abstract protocols --- as presented here --- but concrete implementations as well.

\vfill

\end{document}